\documentclass[referee]{aa-5-mpa}
\usepackage{graphicx}\usepackage{natbib}
\usepackage{amssymb}
\usepackage{txfonts}
\usepackage{natbib}

\usepackage{natbib}
\bibpunct{(}{)}{;}{a}{}{,}  

\newcommand{\deriv}[2]{\frac{\mathrm{d}#1}{\mathrm{d}#2}}

\newcommand{\Iso}[2]{^{#1}{\rm #2}}

\newcommand{\msun}{{\rm M_\odot}}

\newcommand{\lsun}{\,L_\odot}

\begin{document}

\title{On constructing horizontal branch models}

\author{A.~Serenelli\inst{1,2} \and A.~Weiss\inst{1}}

\institute{Max-Planck-Institut f\"ur Astrophysik,
           Karl-Schwarzschild-Str.~1, 85748 Garching,
           Federal Republic of Germany
           \and
           Institute for Advanced Study, Einstein Drive, Princeton, 
	   NJ~08540, USA
           }

\offprints{A.~Weiss}
\mail{A.~Weiss}

\date{Received; accepted}


\abstract{We investigate different methods used to construct
  (zero-age) horizontal branch models and compare the resulting
  horizontal branch evolution with that of models resulting from 
  the calculation of the complete stellar evolution from the main
  sequence and through the core helium flash. We show that the
  approximate methods may lead to small, but discernible effects, but
  that some methods, which are as simple, reproduce the complete
  evolution very well.
\keywords{Stars: evolution -- Stars: horizontal branch} 
}
\maketitle

\clearpage

\section{Introduction}

The horizontal branch (HB) in the Hertzsprung--Russell--Diagram (HRD)
has a number of important applications. For example, it allows
distance-independent age determinations of globular clusters, when the
difference between the time-dependent turn-off brightness and the
almost time-independent one of the HB is used
\citep[e.g.,][]{sw:97}. On the other hand, its brightness or that of
RR~Lyr stars \citep{feast:97} populating it can be used to determine
distances. The number of stars on the HB relative to the number of
stars on the Red Giant Branch (RGB) above the HB is an indicator for
the helium content of the cluster \citep{bpbc:83,srcp:2004}.

Physically, HB stars are in the core helium burning phase following
the RGB. The ignition of helium burning takes place under degenerate
conditions and is a very energetic and partially dynamic event. Stars
of initial mass $M_i \gtrsim 2.3\msun$ ignite Helium under
non-degenerate conditions and populate the equivalent of the HB, the
so-called clump, which is observationally a region of increased star
density close to or on top of the RGB at the level of the HB.

The distribution of post-flash low-mass stars along a horizontal
sequence can be explained by a constant core mass, which is closely
related to the luminosity of the star, and a spread in envelope and
thus total mass \citep{faulkner:66,dorman:92a}. This determines the effective
temperature, but only weakly the luminosity. In a globular cluster,
due to the fact that all stars have the same age, the initial mass of
the stars now populating the HB must have been the same and the spread
along the HB must be the result of different mass loss histories.

The detailed distribution of HB stars as function of $T_\mathrm{eff}$,
the so-called HB morphology, appears to be determined primarily by the
composition (the ``first 
parameter'') and a ``second parameter'', the nature of which is so far
unknown, although age is considered to be a likely candidate
\citep[see, for example][]{ldz:94,sc:98,catelan:2000}. One
should add that here secondary age effects are thought off, because
age itself is a primary parameter as well, due to the fact that
the turn-off mass is already influencing the HB morphology by the
simple fact that younger populations, i.e.\ more massive stars will
end as more massive, therefore cooler, HB stars.

As mentioned above, the evolution to an HB star goes through the
violent core helium flash. For several decades, only very few
calculations have been able to follow stellar models through this
event. The reason is that typical timesteps decrease to the level of
seconds because of the rapid change in helium luminosity, which may
flare up to $10^{10}\lsun$ for a few
days. Because other parts of the star remain completely unchanged
during this time, the differencing schemes are getting
numerically unstable and do no longer converge. Therefore, full flash
calculations were successful 
in most cases only for higher masses, where degeneracy of the
RGB stellar core is lower and the flash milder \citep[see][\ for the
  pioneering study]{hct:67}. With a few exceptions, complete
stellar evolution tracks into the HB phase were practically absent. 

For this reason most HB models were calculated by starting new
sequences on the HB, where the initial structure is taken from that at
the tip of the giant branch, and solutions of the structure equations
are sought where the effect of the helium flash is taken into
account. This implies a reduced core density, hydrostatic helium
burning, a lower surface luminosity and an effective temperature
depending on the envelope mass. Details of such approximate methods
will be given later in the paper. Overall, such starting models
represent the true zero-age HB (ZAHB) and the following evolution
quite well, but there is no systematic investigation about the effects
of the approximation. The underlying assumption of such methods is
that during the helium flash no events take place that modify the
internal chemical structure significantly.

Since a few years an increasing number of stellar evolution codes have
been improved numerically to a level that full flash calculations have
become possible. Still, the computational effort is large, such that
grids of HB models derived from complete evolution sequences are not
truely practicable. However, it is now possible to compare full with
approximate HB models and to design schemes for the construction of HB
models which are both economical and accurate. Given the fact that the
requirements towards the accuracy of stellar models is steadily
increasing, either from observational data about stars, or --- as in the
case of the cluster ages --- because of more detailed questions to be
answered, it appears to be necessary to consider the accuracy of HB
models.

Recently, \cite{ptc:2004} published a similar and independent
investigation testing their own method for ZAHB model
construction. They showed results for two cases, to which we will
refer at the appropriate places.

In the next section we will start this paper with discussing briefly
our stellar evolution code and with presenting full core helium flash
calculations and the resulting ZAHB models and HB evolution. In Sect.~3 we 
present several typical schemes to create approximate HB models without
following the complete flash. Section~4 is devoted to critically analyze results
for approximate models by using results of Sect.~2 as a reference.
We will not investigate methods that rely on a completely new
construction of models, since today they are of only historical
interest \citep{rood:70,sg:76}.
All calculations have been done with the same code to avoid
differences resulting from details of the numerical programs.
Sect.~5 will then close with a short summary.

\section{Complete evolution from the ZAMS to the HB}
\label{sec:2}

\subsection{Code}

Throughout this paper we use the Garching stellar evolution code,
which has been described in some detail in \cite{wsch:2000}. The
numerical aspects of this code have been improved such that the
evolution of Asymptotic Giant Branch (AGB) stars can be followed through
many thermal pulses without convergence problems and intervention
\citep{ww:94a}. Due to the same improvements in the algorithms
controlling the temporal and spatial discretization also the helium
core flash became feasible \citep{josefdiss:96}, because the numerical
problems encountered in both phases are of similar nature. In fact, it
is now possible to calculate through the helium flash even if mixing
of protons from the hydrogen layers into the helium burning region
occurs, which leads to an additional flash event due to the energy
release of carbon conversion to nitrogen. We have presented such
calculations both for Pop.~III \citep{scsw:2001} and Pop.~II models
\citep{cssw:2003}. In the latter case, such mixing occurs when the
helium flash sets in only after the star has already left the RGB due to high
mass loss. For these mixing events, 
(convective) mixing and nuclear burning of both H and He have to be
treated simultaneously because of comparable timescales. In the
present paper, such cases will not be 
considered, because they obviously violate the basic assumption of the
approximate models of an unmodified composition profile between the onset
of the flash and the ZAHB. 

We note that convection is always treated in the most simple way,
i.e.\ using mixing length theory and employing the Schwarzschild
criterion for convective stability. We thus consider neither
overshooting nor semiconvection. Both effects play a role during the HB
evolution, but need additional assumptions and introduce free parameters,
which usually are determined from comparison with observations. We
think that the inclusion would not change our results significantly,
because we are mainly interested in the quality of the constructed
ZAHB models.

\subsection{Calculations} 

\begin{table}
\caption{List of models for which we did full evolutions through the
helium flash. The columns list metallicity, initial mass, the
parameter $\eta$ in Reimers' mass loss formula, 
total mass $M_f$ after the core helium flash, 
core mass $M_c$, \ and convective 
core mass $M_{cc}$ \ at the ZAHB.} 
\label{t:1} 
\begin{center}
\begin{tabular}{cccccc}
\hline $Z$ & $M/M_\odot$ & $\eta$ & $M_f/M_\odot$ & $M_c/M_\odot$ &
$M_{cc}/M_\odot$\\ 
\noalign{\smallskip\hrule\smallskip} 
0.0001 & 0.80 & 0.0 & 0.7971 & 0.5011 & 0.131 \\
            & & 0.3 & 0.6788 & 0.5014 & 0.134 \\
            & & 0.5 & 0.5863 & 0.5015 & 0.133 \\
       & 0.85 & 0.0 & 0.8471 & 0.4996 & 0.132 \\ 
            & & 0.5 & 0.6574 & 0.4999 & 0.131 \\
            & & 0.7 & 0.5632 & 0.5001 & 0.132 \\
0.003 & 0.82 & 0.0 & 0.8170 & 0.4892 & 0.121 \\
           & & 0.2 & 0.6650 & 0.4895 & 0.121 \\ 
           & & 0.3 & 0.5734 & 0.4897 & 0.122 \\
      & 0.90 & 0.0 & 0.8970 & 0.4878 & 0.121\\
           & & 0.3 & 0.6906 & 0.4879 & 0.120 \\
           & & 0.5 & 0.5085 & 0.4880 & 0.119 \\
\noalign{\smallskip\hrule\smallskip}
\end{tabular}
\end{center}
\end{table}

\begin{figure}
\centerline{\includegraphics[draft=false,scale=0.50]{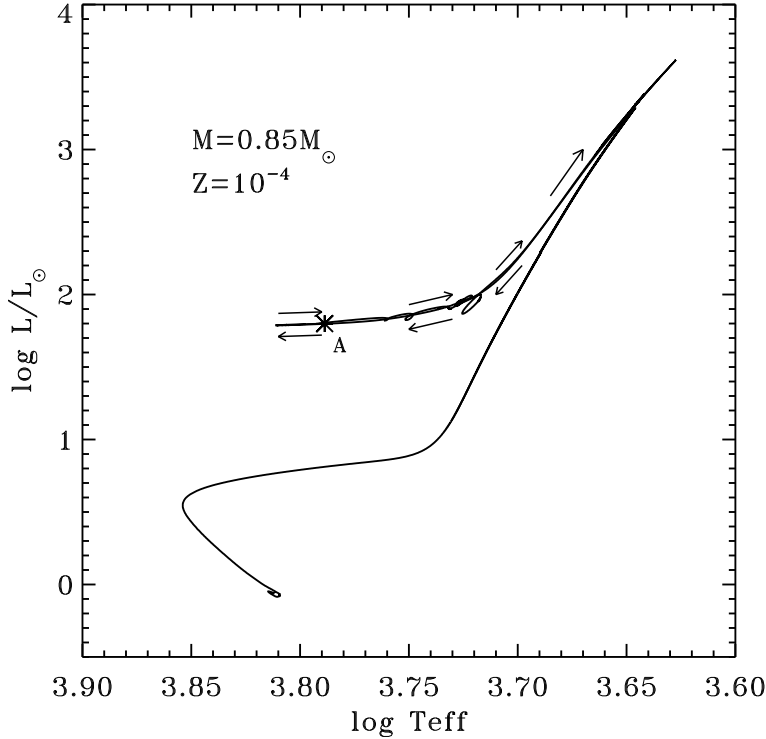}}
\caption[]{Global evolution of one of our models
  from the ZAMS into the early AGB. The sense of evolution from the RGB-tip
  through the HB and up to the  early AGB is indicated by arrows. The
  location of the ZAHB corresponds to the point labeled {\it A}. Note
  that HB evolution is   initially bluewards, as shown by the arrows.}
\protect\label{f:1}
\end{figure}

For this project, we have evolved low-mass stars between $0.8$ and
$0.9\,\msun$, and compositions of  $Z=0.0001$ and $0.003$. The initial
helium content was always $Y=0.25$. A list of cases
is given in Table~\ref{t:1}. 
With these parameters we cover turn-off ages between 8 and 12 Gyr
years and the lower metallicity range of galactic globular
clusters. The evolution was followed from the zero-age main-sequence
(ZAMS), starting from a 
homogeneous model up to the lower AGB, that is prior to the onset of
thermal pulses. For mass loss we used the standard Reimers formula \citep{reimers},
\[
\deriv{M}{t}    =   -4\cdot   10^{-13}    \eta   \frac{LR}{M}    {\rm   \left(
  \frac{M_\odot}{L_\odot R_\odot}\right) }\; {\rm M_\odot}/\mathrm{yr},
\label{e:rml}
\]
varying $\eta$ between $0.0$ (no mass loss) and 
$0.7$ (see Table~\ref{t:1}). The two bluest HB models we have computed are those 
with $Z=0.003$, $0.9\,\msun$, $\eta=0.5$ and $Z=0.0001$, $0.85\,\msun$, 
$\eta=0.7$ which have, on the ZAHB, $T_\mathrm{eff} = 21400$~K and 17400~K 
respectively. 
Our range of parameters resulted from our intention to cover a
wide range of final HB configurations. In Fig.~\ref{f:1} we show the
evolutionary track of the model with initial and constant mass
$M=0.85\,\msun$ and metallicity $Z=0.0001$ (row 4 of Table~\ref{t:1}),
which we will use as a typical case in the discussion following below.

In the calculation of the core helium flash we assume hydrostatic
equilibrium. This is certainly not correct, as our tests show. We
calculated the acceleration $\partial{^2r}/\partial{t^2}$ of mass shells {\em a
posteriori} and compared it with the gravitational
acceleration. Indeed, the two quantities can become of similar
order for a short period after the helium luminosity peak. While our
code allows the inclusion of this approximate 
dynamical term in the hydrostatic equation we have not made use of
this feature.

The question, whether the core helium flash is dynamic and in
particular, whether dynamical episodes are of any consequences for the
flash itself and the later evolution is still unclear. Early
multi-dimensional calculations by Deupree
\citep{dc:83,deupree:84a,deupree:84b} remained inconclusive, but more
recent hydrodynamical calculations (Achatz, M\"uller \& Weiss 2005, in
preparation) indicate that the hydrostatic calculations are indeed
quite well justified, at least from the point of long-term
evolution. In the quoted paper neither any loss of envelope mass nor
any mixing beyond that predicted by mixing length theory of convection
is found. A second argument in favor of the appropriateness of the
hydrostatic models is the fact that so far there is no observational
indication of unknown effects on the HB, which would point to
deviations of the flash evolution from the simple hydrostatic
picture. Nevertheless, the issue remains open, but is of no influence
for the present investigation, which deals only with the influence of
the technical construction of HB models for given physical
assumptions.

\begin{figure}
\centerline{\includegraphics[draft=false,scale=0.50]{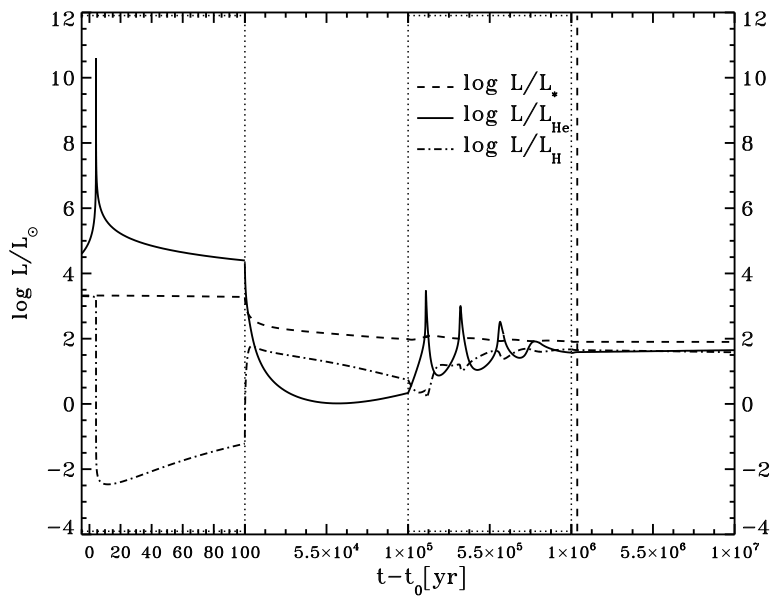}}
\caption[]{Evolution of the total, H- and He-burning 
luminosities for the model shown in Fig.~\ref{f:1} from the onset of 
the He-core flash until the model has  settled on the HB. The moment
when the ZAHB is reached is indicated with a vertical dashed line and  
corresponds to $1.325\times 10^6$~yr after the onset of the flash.} 
\protect\label{f:2}
\end{figure}

In
Fig.~\ref{f:2} we illustrate the evolution of various quantities of the 
model of
Fig.~\ref{f:1} during the helium flash, demonstrating in particular
the extremely short timescales and the amplitude of the helium
luminosity $L_\mathrm{He}$. The evolution during the beginning of the HB is 
also shown in Fig.~\ref{f:2}. While the evolution to the onset of the
flash (defined as the moment in which log~$L_{\rm He}/{\rm L_\odot}$=5 for 
the first time) requires about $9000-13000$ timesteps (from the ZAMS)
depending on the  metallicity and mass of the model,  during the flash another
1500 models are needed, with a minimum timestep of about 600 seconds. 

\begin{figure}[ht]
\centerline{\includegraphics[draft=false,scale=0.45]{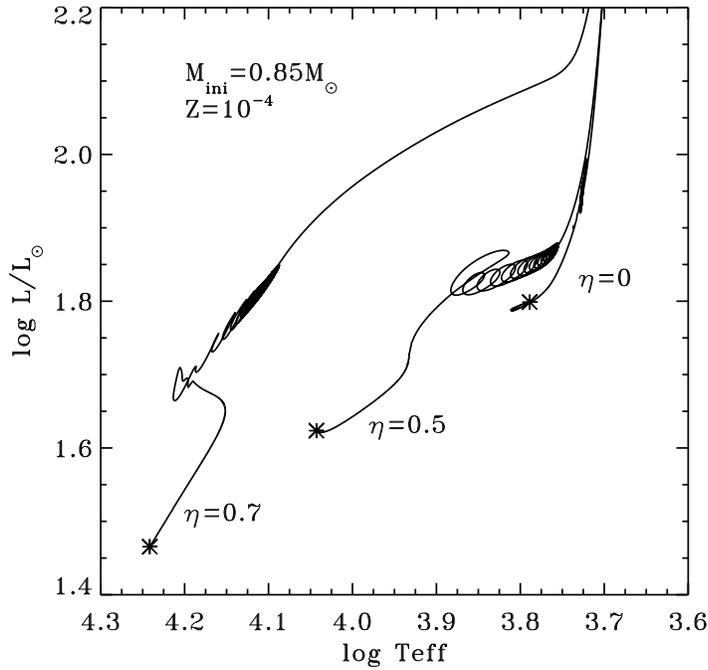}}
\caption[]{Horizontal branch evolution for models evolved from the
  main sequence but with different Reimers mass loss parameter $\eta$
  as indicated in the figure. Initial mass and composition are the
  same as  for the model of  Fig.~\ref{f:1}. We show only the part
  starting at the ZAHB, which is indicated by a star.} 
\protect\label{f:3}
\end{figure}

After the core He flash (and a series of small secondary flashes, as
can be seen in Fig.~\ref{f:1}) convection reaches the center of the
star. At that time the central carbon abundance has already risen to
$X_C=0.0428$ (from an initial value of $1\cdot 10^{-5}$) due to the
initial $3\alpha$-reactions. We define the zero-age on the HB (ZAHB)
to be the model in which the integrated thermal energy is smallest.
For the case shown this happens $1.325\times 10^6$ years after the
onset of the flash (vertical dashed line in Fig.~\ref{f:2} and
asterisk in Fig.~\ref{f:1}) or, equivalently, $0.4906\times 10^6$
years after convection has reached the center.  The central carbon
content when the model reaches the ZAHB is $X_C$=0.0535. This value
depends only slightly on the mass of the models and on the assumed
metallicity (e.g. $X_C=0.05$ for models with $Z=0.003$). This
calculation has been performed without mass loss, therefore resulting
in the reddest ZAHB model possible for the chosen mass and
composition.  In Fig.~\ref{f:3} we display the HB evolution of a
number of cases all evolved from the same initial model, but with
different mass loss rates during their pre-HB evolution as shown in
the diagram (see Table~\ref{t:1}).  There is no additional or enhanced
mass loss during the helium flash except due to changes in the global
stellar parameters entering the Reimers' formula.  For simplification,
we will call HB models obtained from complete evolutionary
calculations simply ``full models''. Mass loss has been halted after
the models reach the ZAHB to facilitate comparisons between ``full
models'' and ``constructed models'' in the following sections.

Stellar evolution during helium-burning phases has been the subject of
numerous studies and it is beyond the scope of this article to review it
here.  One feature, however, is worth commenting on.  We find that
all of our ``full models'' undergo, during the transition from 
helium-core to helium-shell burning phase, a series of small
thermal runaways in the helium-shell.  These pulses, previously found
by \citet{ma:1986} and \citet{alt:2002} in the context of the
evolution of a 3~M$_\odot$ stellar model, have been named thermal
micropulses because of their resemblance to the thermal pulses of AGB
stars.  In the HRD, the micropulses are seen as a series of small
amplitude loops (see Fig.~\ref{f:3}). We find that the occurrence of
these pulses is missed, in most cases, if a large enough
evolutionary time-step is used in the calculations.

\newpage

\section{Construction methods}\label{sec:methods}

In this section we will present a few methods used by various stellar
evolution groups
to  construct (ZA)HB  models,  generally starting  from  RGB models 
evolved into the helium flash.

\subsection{Method 1: Reducing mass along the ZAHB}

This method, named M1 hereafter, is our method to construct
approximate models and differs from others we will describe in as
much as it does not start with an RGB, but rather with a ZAHB model
obtained from a full evolution without mass loss as the one of
Fig.~\ref{f:1}. Once this star settles on the HB and the ZAHB model is
identified, the temporal evolution is stopped, and instead mass is
removed from the envelope until the total mass desired is reached.
The static models obtained after the removal of mass are considered as
true ZAHB stars and their evolution on the HB is started without any
additional relaxation period.  The assumption here is therefore that
all ZAHB models originating from the same initial ZAMS model have the
same chemical structure {\em after the flash} independent of their
mass loss history. This is confirmed by the full model calculations
presented in Sect.~\ref{sec:2}. In Table~\ref{t:1} the size of the
convective core when the models reach the ZAHB is shown to be
identical at a level of $10^{-3}\,\msun$ irrespective of the mass loss
history of the model (it also seems to be quite independent of the
initial mass of the model, at least in the range we have studied). The
same convergence for different models is observed, as stated
previously, in the chemical composition of the core at the ZAHB.

\subsection{Method 2: Reconfiguration of pre-flash models \label{m2}}

This is the standard method for constructing ZAHB models, and has been
used by many groups. Relevant references are 
\cite{ccp:89} and \cite{dlv:91}. In this method a model close to or
at the beginning of the core helium flash is used as the starting
model for a ZAHB construction. The flash itself is not calculated.  In
particular, core mass and chemical profile of the pre-flash model are
kept constant.  The total mass is either preserved, too, or reduced by
``rescaling'' the envelope \citep{dlv:91}.  The new model on the ZAHB
is found by a Runge-Kutta or similar integration with two boundary
conditions each at both center and photosphere. The boundary values
are initially guessed from known HB structures and modified until the
two integrations agree at a predefined fitting point. This new model
is then ``relaxed'' for a certain amount of time or number of
timesteps (depending on the implementation) and afterwards identified
as the new ZAHB model.  

A variant of this method has been used by \cite[and private
communication, 2005]{sweigart:87} by constructing new ZAHB starting models
using homology and other scaling relations and converging these models
immediately with the usual Henyey-solver of the stellar evolution
code.

Already from the first calculations
\citep[see][]{sweigart:87} it was realized that the core composition
on the ZAHB will no longer be the one before the flash, because of a
modest carbon production during the flash. The initial composition for
the new ZAHB model is therefore modified accordingly by hand. The
amount of produced carbon added to the core composition varies between
2 and 5\% by mass, and can be guided by the simple estimate that the
energy needed for the expansion of the degenerate helium core must
come from helium burning \citep{ir:70}.  In the present study, for
which we have full models available, we generally used a value of 5\%
in our calculations, following the results of the calculations
presented in Sec.~\ref{sec:2}. We also tested 2\% and 3\% to
investigate the importance of this guess.

\cite{ptc:2004} investigated the accuracy of ``pseudo-evolutionary''
ZAHB models generated in that way. The relaxation period for these
models is 1 Myr to allow adjustment of CNO equilibrium to the new
thermal structure in the shell region; the additional carbon abundance
is 5\%.  In that paper comparisons with full models are shown and
discussed briefly for $M=0.9$ and $0.794\,\msun$ ($Z=0.0001$;
$Y=0.23$).  The latter case is comparable to our first one of
Table~\ref{t:1}. Note that the core mass in our case is $M_c =
0.5011\,\msun$, while it is $0.5185\,\msun$ in that of
\cite{ptc:2004}.  This reflects the different initial helium content
as well as differences in the treatment of physical aspects like
neutrino cooling or electron conduction.

The details of our implementation of this method are as follows.  We
evolve RGB models until the onset of the flash (log~$L_{\rm He}/{\rm
L_\odot}$=5). We then keep total mass and chemical profile (i.e. the
values of core and envelope mass) fixed, construct a static HB model
using the Runge-Kutta integrator, which is an integral part of our
stellar evolution code and then add 5\% of carbon to the homogeneous
helium core. In the full models, the time between the onset of the
He-flash and the model reaching the ZAHB is $\sim 1.3\times 10^6$
years (a value which depends little on the metallicity and mass of the
model) and during this time some hydrogen burning in the shell also
occurs (basically $\Iso{12}{C}$ is transformed to $\Iso{14}{N}$ in the
layers surrounding the H-burning shell as they get hotter during the
contraction phase from the tip of the RGB to the HB). To account for
this effect, we let the model relax the CNO elements in the H-shell of
the initial HB model for 1.5~Myr while keeping the He content of the
core fixed (it is not a crucial point whether relaxation is allowed
for 1.5 or 1.3~Myr because in both cases CNO abundances rearrange
conveniently; note that also the consumption of protons is taken
into account).  The resulting model is assumed to be the new ZAHB
model. In method M1 this is not necessary, because the method relies
on a full model. It should be emphasized here that in order to minimize the
number of necessary assumptions when using this method we have not
applied any envelope rescaling, i.e. for each HB model we have evolved
accordingly a model from the ZAMS to the RGB-tip with the appropriate
mass loss rate and then relaxed it with the Runge-Kutta integrator.
We call this method M2.

\subsection{Method 3: Rescaling a single initial model} 

This method (M3) is a derivation of the method M2 presented in the
 previous section. First, a HB model corresponding to a stellar model
 evolved from the ZAMS without mass loss is constructed with method M2
 and the ZAHB model is identified. This gives the reddest possible
 ZAHB corresponding to a given ZAMS model. A sequence of ZAHB models
 with smaller envelopes can now be constructed by removing mass from
 the envelope until the desired total mass is reached (in this point
 it resembles method M1).  Numerically, it represents a simplification
 with respect to methods M1 and M2. In the first case, because no full
 flash calculation is needed, and in the second case because only
 one RGB evolution is needed and, at least our personal experience
 points into that direction, it is usually easier to remove mass than to
 relax an RGB model with a Runge-Kutta integrator.

\subsection{Method 4: Relaxing a step-like chemical profile}

One further  simplification that can be done to generate HB  models, which is
equivalent to create them from scratch,  is to avoid the calculation of an RGB
model at all. This can be easily done  by assuming a total and a core mass for
the model,  and giving the desired  composition for the core  and the envelope
assuming  a step-like  shape  at the  core  mass. Because  the required  input
information  for a  Runge-Kutta relaxation  is  the chemical  profile and  the
boundary conditions, HB models can  be generated in this way without involving
the calculation of any previous stage of the evolution of the models. Once the
initial  HB models  has  computed,  CNO abundances  are  relaxed as  described
in Sect.~\ref{m2}. This is method M4.

\section{Results and Discussion} \label{sec:4}

We now compare  the HB  models  constructed with  the approximate  methods
described  in   the  previous  section  with the  full  models   presented  in
Sect.~\ref{sec:2}. We find that  the results of comparing approximate models
to full models can be classified in terms of the location of the models on the
HB, i.e.   in terms  of their  effective temperature. This  is related  to the
morphology  of the evolutionary  track on  the HB  which, as  can be  seen in
Fig.~\ref{f:2}, depends on the effective temperature of the models. 
In this way, we group our
models into  cool ($\log {\rm  T_{eff}} \leq 3.85$), intermediate  ($3.85 \leq
\log 
{\rm T_{eff}} \leq 4.10$) and hot  models ($\log {\rm T_{eff}} \geq 4.1$) and
discuss our results in terms of this classification.

\begin{figure}[t]
\centerline{\includegraphics[draft=false,scale=0.75]{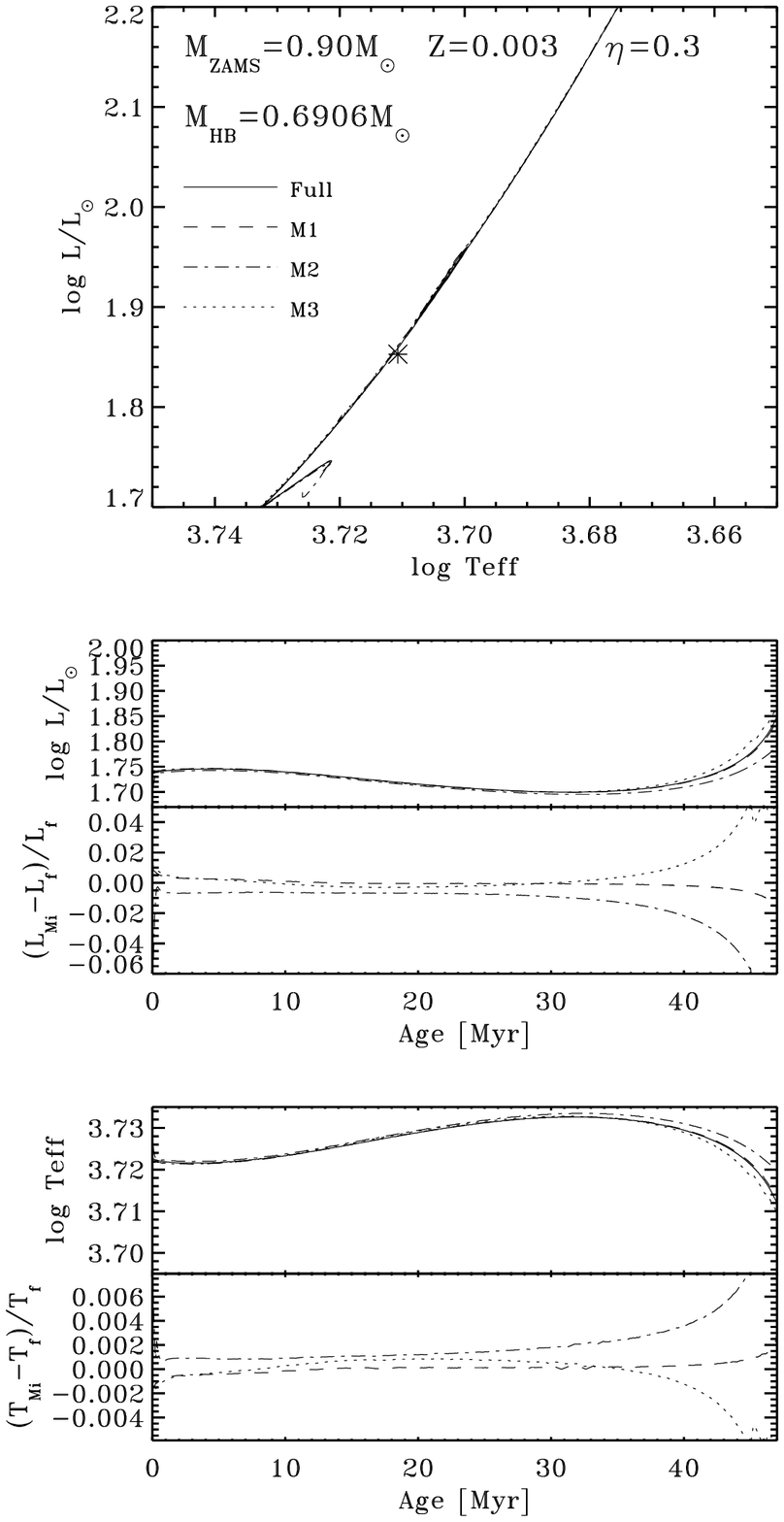}}
\caption[] {Comparison between full and approximate models
constructed with methods M1, M2, and M3 for a typical cool (red) HB
stellar model.  Top panel: evolution in the HRD, starting with the
ZAHB and up to the early AGB. The characteristics of the progenitor
and the mass of the model on the HB are indicated, as well as the
meaning of the different linetypes. The asterisk denotes the point
where the central helium mass fraction drops below 0.01 in the full
model (in the scale of this plot, however, equivalent points in all
sequences overlap almost completely).  Middle panels: luminosity and
relative luminosity differences for approximate models with respect
to the full model against time. Indices $f$ and $Mi$ indicate full
models respectively the i-th method.  Bottom panels: effective
temperature and relative effective temperature differences against
time.  Ages are counted from the ZAHB. Middle and lower panels show
the evolution until the end of the HB evolution only, defined by the
end of core helium burning.} 
\protect\label{f:4}
\end{figure}

Let us start by making a short comment on method M4.  We have employed
this method to construct models corresponding to the $M=0.80\,\msun$,
$\eta=0.3$ and $M=0.82\,\msun$, $\eta=0.2$ full models.  The
difference in effective temperature and luminosity between the models
constructed with this method and full models is approximately 5\%
along the whole HB evolution. This level of agreement, although
reasonably good considering that the models are basically constructed
from scratch, is at least a factor of two worse than that obtained
with any of the other methods presented in this paper as will be shown
below.  For this reason, no further tests have been carried out with
method M4 and no more comments are devoted to it. We focus our
discussion on methods M1, M2 and M3, which have been used to obtain HB
models for all relevant cases presented in Table~\ref{t:1}.

\begin{figure}
\centerline{\includegraphics[draft=false,scale=0.75]{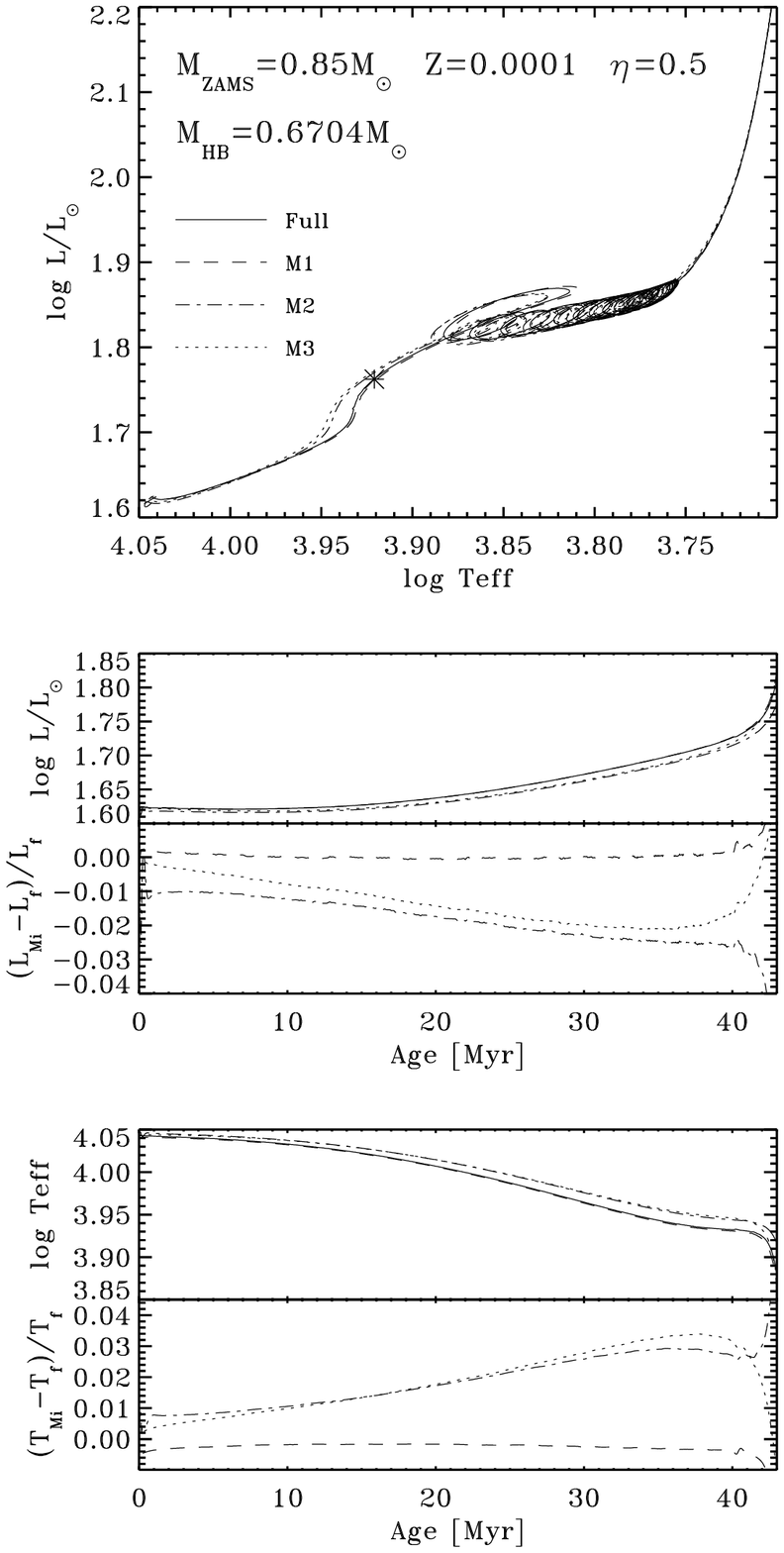}}
\caption[]{The same as Fig.~\ref{f:4} but for a typical intermediate HB
model. Characteristics of the model are given in the top panel.}
\protect\label{f:5}
\end{figure}

\begin{figure}
\centerline{\includegraphics[draft=false,scale=0.75]{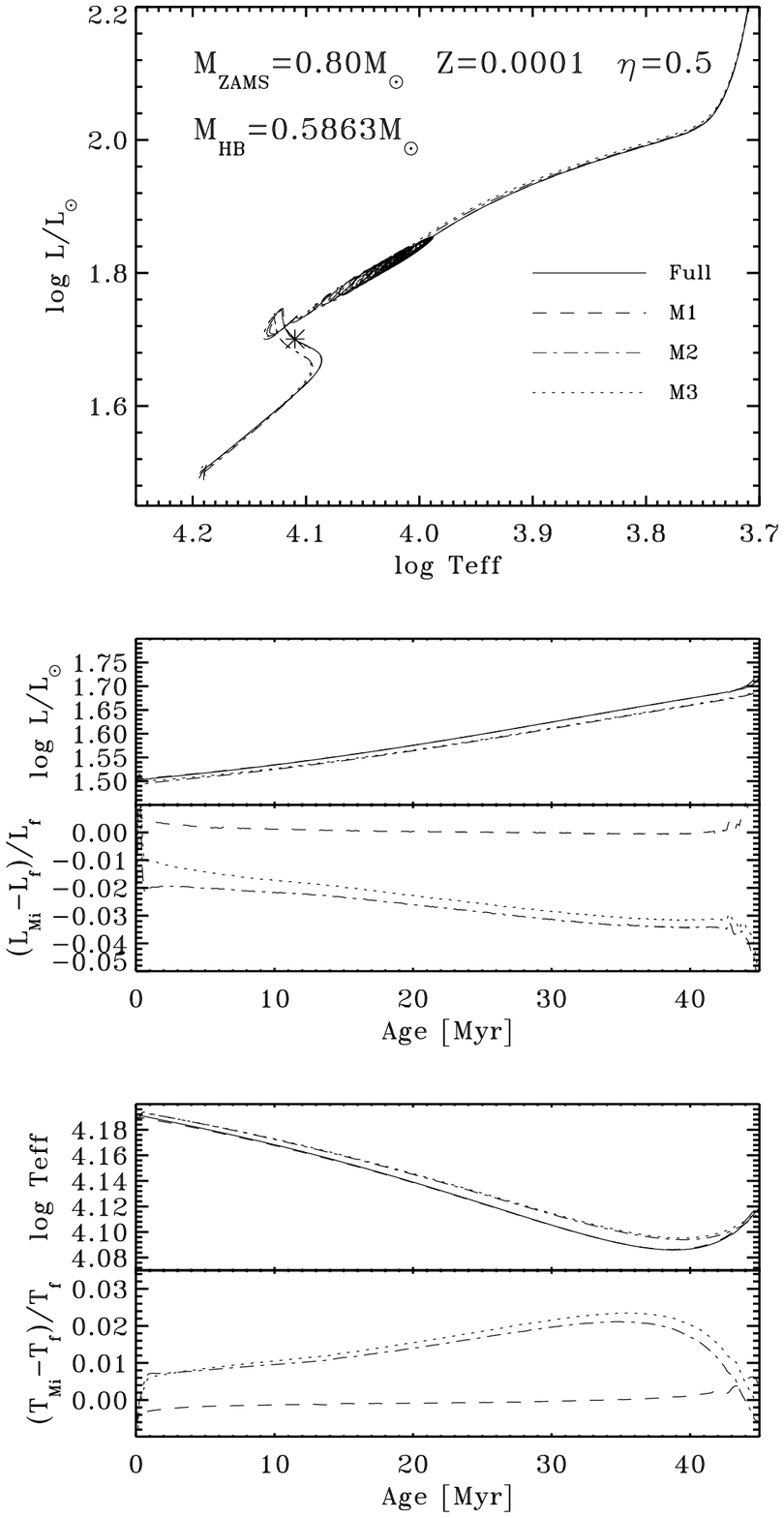}}
\caption[]{The same as Fig.~\ref{f:4} but for a typical hot HB
model. Characteristics of the model are given in the top panel.}
\protect\label{f:6}
\end{figure}

In Fig.~\ref{f:4} we present the results for a typical {\em cool} HB
model ($Z=0.003$, $M= 0.6906\,\msun$). The top panel shows the
evolution of the model in the HRD. Results for the M1, M2 and M3
methods are also shown but are almost indistinguishable in this
plot. Differences in luminosity and effective temperature with respect
to the full model as function of time are also displayed in the plot.
The two panels of the central plot show the behavior of luminosity
against time during HB evolution for the full and M1, M2, and M3
models, and the relative difference in luminosity for the
approximate models with respect to the full one. All methods give
good agreement with predictions of the full model for most of the HB
evolution and curves practically overlap in the luminosity
plot. However, it can be seen that in the last stages of the HB
evolution (between 35 and 45 Myr), M2 and M3 models deviate from the
full model, with differences that reach up to 4\% for M3 model and
more than 5\% in the case of method M2.  The reason for the difference
is that for the M2 model, evolution is slightly slower than in the
full model, while for M3 the situation is reversed.  Deviations are
more noticeable at the last stages of HB evolution because at those
stages luminosity changes occur on much shorter timescales than in
previous stages. The M1 model completely overlaps the full model in
luminosity plot and this can also be seen as a constant difference of
about 0.5\% with respect to the full model during the whole HB
evolution.  The bottom panels show the behavior of effective
temperature against time in the same fashion as for luminosity. The
conclusions are qualitatively the same. The relative differences
between M2 and M3 with respect to the full model are much smaller in
magnitude and the sign is the opposite of that of the luminosity
differences.  Again, method M1 yields the most accurate
agreement with the full models. In summarizing these results it may be
interesting to note that, although all full and approximate models
basically move along the same evolutionary track as seen in the HRD,
they do it at a somewhat different speed so that at a given age they
are in slightly different position on it. This may have some influence
on applications based on star counting on the HB, although the effect
will be rather small, since the age differences generally remain below
1~Myr.

We discuss now our results for {\em intermediate } HB models,
i.e. those with effective temperatures around 10000~K. As a
representative of this group we show the model with $Z=0.0001$ and $M=
0.6704\,\msun$, as indicated in the top panel of Fig.~\ref{f:5}, where
the HRD evolution of the full and approximate models are also
presented.  In this case, the HRD diagram reveals some differences
between the full and corresponding M2 and M3 models, specially
noticeable at about $\log T_{\rm eff} \approx 3.94$. The M1 model, on
the other hand, completely overlaps the full model for the whole HB
evolution (which ends at the point marked with an asterisk). In the
middle and bottom panels it is clearly seen that the M1 model gives
also a very good agreement with the full model for the temporal
evolution of luminosity and effective temperature (at the level of
0.4\% or less).  M2 and M3 models have very similar behavior, being
that the deviations with respect to the full model increase
monotonically during almost all the HB evolution and reach values that
vary between 2 and 4\%.

The case for a {\em hot} HB models is presented in Fig.~\ref{f:6},
where results for the model $Z=0.0001$ and $M= 0.5863\,\msun$ are
shown. M1 model performs very well in this case, too, at the same
level as in the preceding cases. It cannot be distinguished from the
full model in the HRD and in the plots where luminosity and effective
temperature are given as a function of time. Models constructed with
methods M2 and M3 lead to a slightly different HRD evolution than the
full or M1 models.  Also, as in the case for intermediate HB models,
the difference between these models and the full model tends to
increase with time. Notice that the lower panels only cover the
evolution up to the end of the HB; after that, during the micropulses,
the deviations become  much larger because of the  rapid changes in luminosity
and effective temperature (Fig.~\ref{f:6},  upper panel). On average, however,
good agreement is still present. 
 
Let us comment on the results for the most extreme HB model we have
calculated ($Z=0.003$, $0.9\,\msun$, $\eta=0.5$), that has $\log
{T_{\rm eff}}\approx 4.32$.  The location on the HRD of such extreme
models, for a given stellar mass, is extremely dependent on the
hydrogen-depleted core mass (or, for given core mass, on the envelope
mass). The full model has total mass $M_\mathrm{HB}= 0.5085M_\odot$ and
core mass $M_c=0.4880M_\odot$.  The corresponding M1 model, constructed
from the $Z=0.003$, $0.9\,\msun$, $\eta=0$ model has the same total
mass by construction but its hydrogen-depleted core mass is
$M_c=0.4878M_\odot$.  This very small (0.04\%) difference in the core
size translates into 1\% in the envelope mass. As a result, the
M1 model effective temperature and luminosity along the HB differ by
about 1\% and 0.5\% from those of the full model respectively.

There is another complication for even higher mass loss and thus
even more extreme HB models. Above the hydrogen burning shell there is
always a region of varying CNO-abundances (mostly C and O are reduced)
due to partial
CNO-cycling. With increasing mass loss on the RGB this region is
squeezed closer to the shell. However, even for method M1, our
approach rests on the assumption that the chemical profile is
independent of the mass loss history, which in this extreme case, is
no longer justified. In practice, if we were to construct ZAHB models
with total mass of order $0.55\,\msun$ or below, we would remove part
of the region of varying composition, uncovering processed layers, in
contrast to full models. While there might not be a significant change
in the following HB evolution, the surface composition will be
incorrect. We have therefore avoided such extreme models and recommed
full models in such cases. Sweigart (2005, private communication) has
been using a method similar to M1 which considers above mentioned
effect as well by reducing the mass of the C- and O-depleted regions.

Methods M2 and M3 provide the freedom of choosing the central composition
of the ZAHB model, i.e. how much helium is assumed to have been converted
into carbon during the flash. Based on full calculations, our standard
choice for  M2 and  M3 models has  been that 5\%  of the  mass in the  core is
carbon at the 
ZAHB. We have tried, however, some models where this value was set to
2 and 3\%.  The first choice does not give a good agreement with full
models.  If 3\% of carbon in the core is assumed, then the
quantitative agreement with full models is similar to that of M2 and
M3 models with 5\% initial carbon content, the discrepancies being of
opposite signs. This seems to point in the direction that 4\% of
central carbon abundance is the best choice for M2 and M3 methods, a
value that is, however, not the one indicated by full models and
implies that M2 and M3 models may require fine tuning to reproduce
results as good as those of M1 method.

Finally,  we  have  computed a  few  cases  where  full models  were  computed
including microscopic  diffusion from the ZAMS.  We have tested  our M1 method
against these models and found the same performance as described above 
for models without  diffusion.  This is an expected  result, basically because
method M1 incorporates, by construction, all the hypotheses used in the 
computation of the full model from which M1 models are constructed.

\section{Summary and Conclusions}

The aim of the paper is to analyze the accuracy of HB models
constructed with a variety of approximate methods by comparing them
to full evolutionary models, i.e. models obtained by following their
evolution from the ZAMS up to the RGB-tip and through the core helium 
flash. In particular, we present our method (M1) which consists in
evolving a given initial ZAMS model without mass loss all the way from
the ZAMS to the ZAHB and then remove envelope mass while stopping the
evolution to get the final HB model mass desired. In this way, a whole
HB sequence can be built with a negligible amount of effort once the
first ZAHB model is given. Other, and maybe more important, advantages
of this method consist in that no guess for the amount of helium
burned during the flash and no relaxation time to adjust CNO
abundances in the hydrogen shell are needed.  These last two points
are potential sources of uncertainties in the methods usually
presented in the literature (see description of methods M2 and M3 in
Sect.~\ref{sec:methods}). Comparison with the full models shows that
any of the three methods discussed in this paper give reliable HB
models, but the performance of method M1 is outstanding. This is, then,
our preferred method.

We confirm the results by \cite{ptc:2004} about their method (M2), but
find that this and method M3 are less accurate in particular for
hotter HB stars. With increasing $T_\mathrm{eff}$ (decreasing envelope
mass) the discrepancies get to a level of a few per cent. Also,
details of the post-HB evolution (micropulses) differ. However, this
phase also depends on the treatment of convection on the HB, such that
we did not consider these events in detail. 

While we have not investigated cases with overshooting or some treatment
of semiconvection, we don't think that our results would change, in
particular because the FRANEC code used in \cite{ptc:2004}
incorporates the latter and our results agree with theirs.

\begin{acknowledgements}
We thank S.~Degl'Innocenti, L.~Girardi, and S.~Cassisi for information
concerning their methods for constructing horizontal branch models.
H.~Schlattl supported us in the calculations of full flash models and
had the original idea for our prefered method M1. We thank in
particular A.~Sweigart for an excellent referee report which contained
very valuable additional information and helped to improve this paper.

AMS has been supported in part by the European Association for Research in 
Astronomy through its EARASTARGAL program, the W. M. Keck Foundation 
through a  grant to the Institute for  Advanced Study and the  NSF through the
grant PHY-0070928. We are grateful for the hospitality we experienced
at our mutual visits at MPA (Garching) and IAS (Princeton).

\end{acknowledgements}

\newpage
\bibliographystyle{aa}
\bibliography{hb}

\
\end{document}